\documentclass[aps,pra,reprint,groupedaddress,showkeys]{revtex4-2}
\usepackage{amsmath,graphicx,hyperref,verbatimbox}
\usepackage{indentfirst}
\usepackage{braket}
\usepackage{diagbox}
\usepackage{amssymb}
\usepackage{bbold}
\usepackage{colortbl}
\usepackage{multirow}
\usepackage{subcaption}

\newcommand{\figref}[1]{Fig.~\ref{#1}}

\renewcommand{\eqref}[1]{Eq.~\ref{#1}}
\newcommand{\tr}{{\mathrm{Tr}}}
\newcommand{\centered}[1]{\begin{tabular}{l} #1 \end{tabular}}

\begin{document}

\title{Entanglement characterization by single-photon counting with random noise}
\author{Artur Czerwinski}
\email{aczerwin@umk.pl}
\affiliation{Institute of Physics, Faculty of Physics, Astronomy and Informatics \\ Nicolaus Copernicus University in Torun,
Grudziadzka 5, 87--100 Torun, Poland}

\begin{abstract}
In this article, we investigate the problem of entanglement characterization with polarization measurements combined with maximum likelihood estimation (MLE). A realistic scenario is considered with measurement results distorted by random experimental errors. In particular, by imposing unitary rotations acting on the measurement operators, we can test the performance of the tomographic technique versus the amount of noise. Then, dark counts are introduced to explore the efficiency of the framework in a multi-dimensional noise scenario. The concurrence is used as a figure of merit to quantify how well entanglement is preserved through noisy measurements. Quantum fidelity is computed to quantify the accuracy of state reconstruction. The results of numerical simulations are depicted on graphs and discussed.
\end{abstract}
\keywords{quantum state tomography, photon counting, entanglement characterization, concurrence}
\maketitle

\section{Introduction}

\textit{Quantum state tomography} (QST) aims at recovering accurate mathematical representations of quantum states from measurements \cite{dariano03,paris04}. Usually, the post-measurement state of the system is of little interest since the probabilities of the respective measurement outcomes are at the center of attention. In such cases, positive operator-valued measures (POVMs) can be applied to study the statistics of measurements \cite{Nielsen2000}. In particular, symmetric informationally complete POVMs (SIC-POVMs) can be considered optimal as far as the number of measurements is concerned \cite{Rehacek2004,Renes2004,Fuchs2017}.

Some proposals, both theoretical and experimental, focus on performing QST with the minimal number of measurements \cite{Oren2017,Martinez2019}. In particular, methods based on compressed sensing can decrease the number of measurement settings \cite{Gross2010}. Special attention should also be paid to the methods that utilize dynamical maps to decrease the number of necessary measurement operators \cite{Merkel2010,Czerwinski2016a,Czerwinski2020a}.

On the other hand, in practical realizations of QST protocols, there is a tendency to apply overcomplete sets of measurements in order to reduce the detrimental impact of experimental noise \cite{Horn2013,Zhu2014}. Particularly, mutually unbiased bases (MUBs) can be employed as an overcomplete measurement scheme \cite{Wootters1989,Durt2010}.

Quantum mechanics admits the existence of physical systems which feature correlations that cannot be explained based on the classical physics \cite{Einstein1935}. In this context, we usually consider compound quantum systems which exhibit nonlocal quantum correlations that can be verified experimentally by detecting multiparticle quantum interference \cite{Bell1964}. In recent years, quantum entanglement, i.e., a specific form of non-classical correlations, has been studied in terms of both creation and detection \cite{Horodecki1996,Horodecki2009}. In particular, bipartite states have been explored by means of different entanglement measures \cite{Schwaiger2018}. 

Entangled states are considered an essential resource in quantum communication and information \cite{Ekert1991}. For this reason, the ability to characterize entanglement based on measurements plays a crucial role in practical realizations of quantum protocols. In the case of photons, quantum information can be encoded by exploiting different degrees of freedom: polarization, spectral, spatial, and temporal mode. Each approach requires distinct measurement schemes for state tomography. In this work, we focus on two-photon polarization-entangled states.

Polarization-entangled photons, produced by a spontaneous-down-conversion photon source, were successfully characterized by polarization measurements \cite{White1999}. Then, photonic state tomography was developed in terms of both theory and experiments \cite{James2001,Altepeter2005,Matsuda2012}. Sources that generate both distinguishable and indistinguishable entangled photon pairs were considered in terms of the visibility of two-photon interference, including all the optical losses and detector's efficiency \cite{Takesue2009}. Since noise and errors are inherent, we implement methods that produce reliable estimates of actual quantum states, such as maximum likelihood estimation (MLE) \cite{Hradil1997,Banaszek1999}, hedged maximum likelihood (HMLE) \cite{BlumeKohout2010a}, the method of least squares (LS) \cite{Opatrny1997} or Bayesian mean estimation (BME) \cite{BlumeKohout2010}. Quantum state estimation methods differ in accuracy, and their efficiency can be compared \cite{Acharya2019}.

In the present article, we consider state estimation of two-qubit entangled states from polarization measurements with photon counts distorted by three types of errors: the Poisson noise, dark counts, and random rotations. The latter can be attributed to errors due to angular setting uncertainties. We analyze selected figures of merit versus the amount of noise.

In Sec.~\ref{methods}, we introduce the framework and all assumptions concerning measurement results. Next, in Sec.~\ref{performance}, we define the figures of merit used to evaluate the accuracy of the model. Then, in Sec.~\ref{results}, the results are depicted on graphs as we compare different scenarios. The findings are analyzed to formulate general conclusions. In the last part, i.e., Sec.~\ref{discussion}, we discuss the results and indicate open problems for future research.

\section{Entangled state tomography with noisy measurements}\label{preliminaries}

\subsection{Methods}\label{methods}

We assume that our source can generate two classes of maximally entangled two-photon quantum states:
\begin{equation}\label{m0}
\ket{\Phi (\alpha)} = \frac{1}{\sqrt{2}} \left(\ket{00} + e^{i \alpha} \ket{11} \right)
\end{equation}
\begin{equation}\label{m1}
\ket{\Psi (\beta)} = \frac{1}{\sqrt{2}} \left(\ket{01} + e^{i \beta} \ket{10} \right),
\end{equation}
where $\{\ket{00}, \ket{01}, \ket{10}, \ket{11}\}$ denotes the standard basis in $\mathcal{H}$ such that $\dim \mathcal{H} =4$. The parameters $\alpha$ and $\beta$ can be considered relative phases between the corresponding basis states. Such state vectors comprise the four Bell states, which are famous for multiple application in quantum communication and information.

The measurement scheme is based on photon-counting in different polarization settings. We assume that the polarization analyzers can be adjusted for six types of polarization: horizontal, vertical, diagonal, anti-diagonal, left-circular, and right-circular. For the tomographic reconstruction of a two-photon entangled state, we consider every combination of these measurements. As a result, we utilize an overcomplete set of $36$ polarization measurements. Let us denote the set of the $4 \times 4$ measurement operators by $M_1, \dots, M_{36}$.

To make the framework realistic, we assume that the measurements are subject to experimental noise, which can be mathematically modeled by random unitary transformations that distort the original measurement operators, cf. Ref.~\cite{Lohani2020,Danaci2021}. The general form of a $2 \times 2$ unitary rotational operator is given by:
\begin{equation}\label{m2}
U (\omega_1, \omega_2, \omega_3) = \begin{pmatrix} e ^{i \omega_1/ 2 } \cos \omega_3 & & -i e^{i \omega_2} \sin \omega_3 \\ & & \\ - i e^{-i \omega_2} \sin \omega_3 & & e^{- i \omega_1/2} \cos \omega_3  \end{pmatrix},
\end{equation}
where the parameters: $\omega_1, \omega_2, \omega_3$, in our application, are selected randomly from a normal distribution characterized by the mean value equal $0$ and a non-zero standard deviation denoted by $\sigma$, i.e. $\omega_1, \omega_2, \omega_3 \in \mathcal{N}(0,\sigma)$. This allows us to construct a $4\times4$ perturbation matrix $\mathcal{P} (\sigma)$ defined as:
\begin{equation}\label{m3}
\mathcal{P} (\sigma) := U (\omega_1, \omega_2, \omega_3) \otimes U (\omega'_1, \omega'_2, \omega'_3).
\end{equation}
The perturbation matrix of the form \eqref{m3} can be attributed to errors due to the apparatus settings, i.e., uncertainties in the settings of the angles of the waveplates used to perform the tomographic projections of quantum states.

Equipped with the definition \eqref{m3}, we can introduce a perturbed measurement operator $\tilde{M}_k (\sigma)$ burdened with experimental uncertainty:
\begin{equation}\label{m4}
\tilde{M}_k (\sigma) := \mathcal{P}_k (\sigma) \, M_k \, \mathcal{P}_k^{\dagger} (\sigma)
\end{equation}
For every act of measurement, a different perturbation matrix $\mathcal{P}_k (\sigma)$ is generated with random parameters according to \eqref{m3} and \eqref{m2}, which allows us to obtain noisy measurement results with a given parameter $\sigma$. Thanks to this approach, every act of observation is burdened with random uncertainty, and $\sigma$ is used to quantify the amount of experimental noise.

Then, if $\rho_{in}$ stands for the input state generated by the source, we obtain a formula for the measured photon count associated with $k-$th measurement operator:
\begin{equation}\label{m14}
n^M_k (\sigma) := \mathcal{N}_k \, \tr \left( \tilde{M}_k (\sigma) \rho_{in} \right),
\end{equation}
where $\mathcal{N}_k$ is generated randomly from the Poisson distribution, i.e. $\mathcal{N}_k \in \mathrm{Pois} (\mathcal{N})$, where $\mathcal{N}$ represents the average number of photon pairs involved in one measurement. The formula \eqref{m14} includes the Poisson noise \cite{Hasinoff2014}, which is considered a standard source of uncertainty in photon counting and has to be taken into account in photonic state tomography \cite{Mohammadi2014,SedziakKacprowicz2020,Czerwinski2021}.

The measurement results defined in \eqref{m14} describe a bulk process, which means that our source can produce a beam consisting of a large number of identically prepared photons. Each photon undergoes the measurement individually, and we aggregate signals to obtain the measured count. The formula \eqref{m14}, which gives the average number of detections, allows us to mathematically model an experimental scenario with random noise.

Then, we can consider an extended noise scenario in such a way that dark counts are incorporated into the model. We assume that, apart from the intended state, the polarization analyzers receive photons from the background. The quantum state of the background noise is described by a maximally mixed state. Thus, the state which undergoes the measurements can be given by:
\begin{equation}\label{m15}
\tilde{\rho}_{in} (p) = (1-p) \rho_{in} + \frac{p}{4}\, \mathbb{1}_4,
\end{equation}
where $\mathbb{1}_4$ represents a $4 \times 4$ identity matrix and $p$ quantifies the amount of dark counts. Then, the number of two-photon counts which reach the detectors can be expressed as:
\begin{equation}\label{m16}
\tilde{n}^M_k  (\sigma, p) = \mathcal{N}_k \, \tr \left( \tilde{M}_k (\sigma) \tilde{\rho}_{in} (p) \right),
\end{equation}
which means that for an arbitrary $\sigma$, $p$ and $\mathcal{N}$ we are able to generate numerically experimental data corresponding to any input state of the form \eqref{m1}.

However, when we reconstruct an unknown state of a quantum system, we assume that the experimenter does not possess any a priori knowledge about the state in question. Thus, we utilize the Cholesky factorization, cf. Ref.~\cite{James2001,Altepeter2005,SedziakKacprowicz2020}, which provides a general representation of a $4\times4$ density matrix:
\begin{equation}\label{m5}
\rho_{out} (t_1, \dots, t_{16}) = \frac{T^{\dagger} T}{\tr\: \left(T^{\dagger} T\right)},
\end{equation}
where:
\begin{equation}\label{m6}
T=\begin{pmatrix} t_1 & 0 & 0 &0 \\ t_5 + i\, t_6 & t_2 & 0 &0 \\  t_{11} + i \,t_{12} & t_7 + i\, t_8 & t_3 &0 \\ t_{15} + i\, t_{16} & t_{13} + i\, t_{14} & t_9 + i\, t_{10} & t_4 \end{pmatrix}.
\end{equation}
This approach means that we need to estimate the values of $16$ real parameters: $\mathcal{T} = \{t_1, t_2, \dots, t_{16}\}$ in order to obtain the complete knowledge about an unknown state. Thanks to the Cholesky decomposition, any density matrix resulting from the framework is physical, i.e., it is Hermitian, positive semi-definite, of trace one.

Based on already introduced symbols, we can write a formula, according to the Born's rule, for the expected two-photon count in the $k-$th measurement:
\begin{equation}\label{m7}
n^{E}_k = \mathcal{N} \, \tr \left( M_k \,\rho_{in} \right).
\end{equation}
In order to determine the values of the parameters $t_1, \dots, t_{16}$ that fit optimally to the noisy measurements, we shall follow the method of maximum likelihood estimation (MLE). We apply the likelihood function $\mathcal{L}$ in the form \cite{Ikuta2017}:
\begin{equation}
\mathcal{L}(\mathcal{T}) = \sum_{k=1}^{36} \left[\frac{(n_k^M -  n_k^E)^2}{ n_k^E }  + \ln n_k^E \right],
\end{equation}
where the measured photon counts, $n_k^M$, are substituted with either \eqref{m14} or \eqref{m16} depending on the noise scenario which is selected for consideration.

\subsection{Performance analysis}\label{performance}

In the present work, we investigate the efficiency of the tomographic framework for polarization entangled states in different noise scenarios. The quality of state reconstruction is quantified by one figure of merit: quantum fidelity, $\mathcal{F} (\sigma)$, given by \cite{Nielsen2000}:
\begin{equation}\label{m9}
\mathcal{F} (\sigma) := \left(\tr \sqrt{\sqrt{\rho_{out}} \, \rho_{in}  \, \sqrt{\rho_{out}}} \right)^2 = \bra{x} \rho_{out} \ket{x},
\end{equation}
where the last formula is due to the fact that the input density matrix is always a pure state, i.e. $\rho_{in}=\ket{x}\!\bra{x}$, where $\ket{x}$ represents either class of the Bell states \eqref{m0}-\ref{m1}. In our framework, the figure \eqref{m9} depends on the amount of noise introduced into the measurements, quantified in general by three figures: $\sigma, p$, and $\mathcal{N}$. We treat $\sigma$ as our independent variable, whereas $p$ and $\mathcal{N}$ are considered parameters. For two quantum states, the fidelity measures their closeness (overlap) \cite{Uhlmann1986,Jozsa1994}. This quantity is commonly used to compare the result of a QST framework $\rho_{out}$ with the original state produced by the source $\rho_{in}$, cf. Ref.~\cite{paris04,Horn2013,Czerwinski2021b}.

In our model, each time we perform QST for a sample of input states defined as \eqref{m0}-\ref{m1}. Every input state $\rho_{in}$ is compared with the result of estimation, $\rho_{out}$, by calculating the fidelity \eqref{m9}. Then, the robustness of the framework against noise can be expressed by the average fidelity $\mathcal{F}_{av} (\sigma)$ computed over the sample. A similar approach to evaluate the performance of quantum state estimation techniques was utilized in Ref.~\cite{SedziakKacprowicz2020,Czerwinski2021}.

Furthermore, in order to quantify how well entanglement is detected by the noisy measurements, we compute for each density matrix (obtained from the QST scheme) its concurrence, which is a convenient entanglement measure for two-qubit states \cite{Hill1997,Wootters1998}. To begin with, for a density matrix $\rho_{out}$, we construct the spin-flipped state, denoted by $\tilde{\rho}_{out}$, which is defined as:
\begin{equation}\label{m11}
\tilde{\rho}_{out} := \left(\sigma_y \otimes \sigma_y \right) \rho^*_{out} \left(\sigma_y \otimes \sigma_y \right),
\end{equation}
where $\rho^*_{out}$ represents the complex conjugate of $\rho_{out}$ (provided one operates in the standard basis) and $\sigma_y$ stands for one of the Pauli matrices, i.e. $\sigma_y = \begin{pmatrix} 0 & -i \\ i & 0 \end{pmatrix}$. Next, we compute the $R-$matrix:
\begin{equation}\label{m12}
R := \sqrt{\sqrt{\rho_{out}} \,\tilde{\rho}_{out}\, \sqrt{\rho_{out}}},
\end{equation}
which finally leads to the definition of the concurrence, $C[\rho_{out}]$, which is expressed by the eigenvalues of the $R-$matrix:
\begin{equation}\label{m13}
C[\rho_{out}] := \max \left\{0, \alpha_1 - \alpha_2 - \alpha_3 -\alpha_4   \right\},
\end{equation}
where $\alpha_1, \alpha_2, \alpha_3, \alpha_4$ denote the eigenvalues of the $R-$matrix arranged in the decreasing order.

The concurrence is a legitimate entanglement measure since, for any density matrix $\rho$, we have: $0\leq C[\rho] \leq 1$, where $C[\rho] = 1$ for a maximally entangled state, and $C[\rho] = 0$ for a separate state. The concurrence is an entanglement monotone, which makes it an appropriate figure of merit to quantify two-qubit entanglement. In particular, it is applied to quantify entanglement preservation subject to noise and inaccuracies, see, e.g., Ref.~\cite{Walborn2006,Buchleitner2007,Neves2007,Bergschneider2019}. Concurrence is directly connected with another fundamental measure, which is called \textit{entanglement of formation} \cite{Bennett1996,Horodecki2001}.

Lastly, to evaluate the performance of the framework for a sample of states, we are using the average concurrence, denoted by $C_{av} (\sigma)$, which is treated as a function of $\sigma$, whereas $p$ and $\mathcal{N}$ are considered parameters. This figure of merit allows us to investigate entanglement detection versus the amount of experimental noise introduced by random rotations.

\section{Results and analysis}\label{results}

\subsection{Entanglement analysis by polarization measurements}

	\begin{figure*}[ht!]
		\centering
		\begin{tabular}{|c|c|c|}
\hline
			\backslashbox[18mm]{$\mathcal{N}$}{$p$}&\centered{$p=0$}&\centered{$p=0.25$}\\\hline
			\centered{$1\,000$}& \begin{tabular}{l}\raisebox{- \height}{\includegraphics[width=0.85\columnwidth]{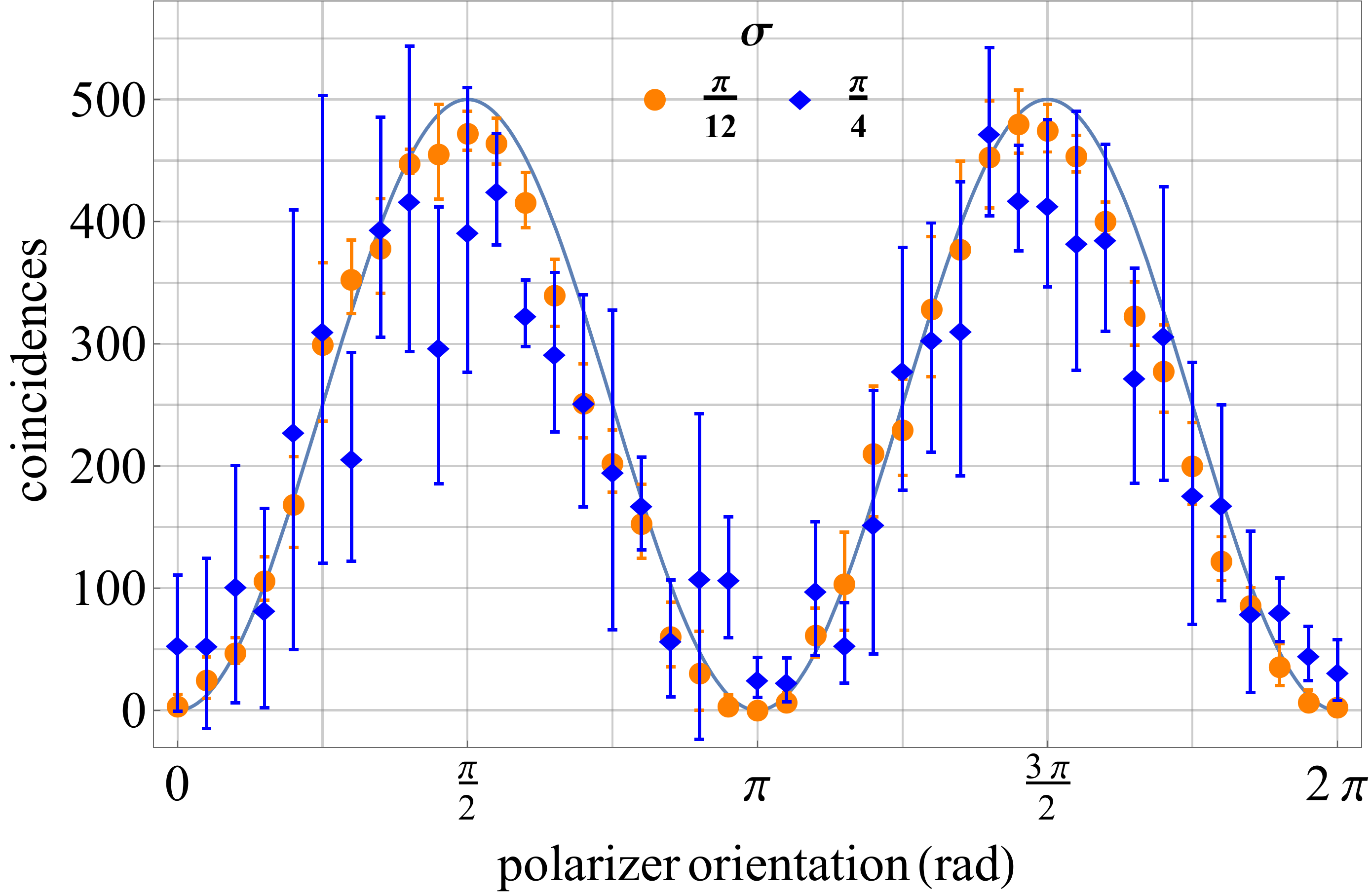}}\end{tabular}&
			\begin{tabular}{l}\raisebox{- \height}{\includegraphics[width=0.85\columnwidth]{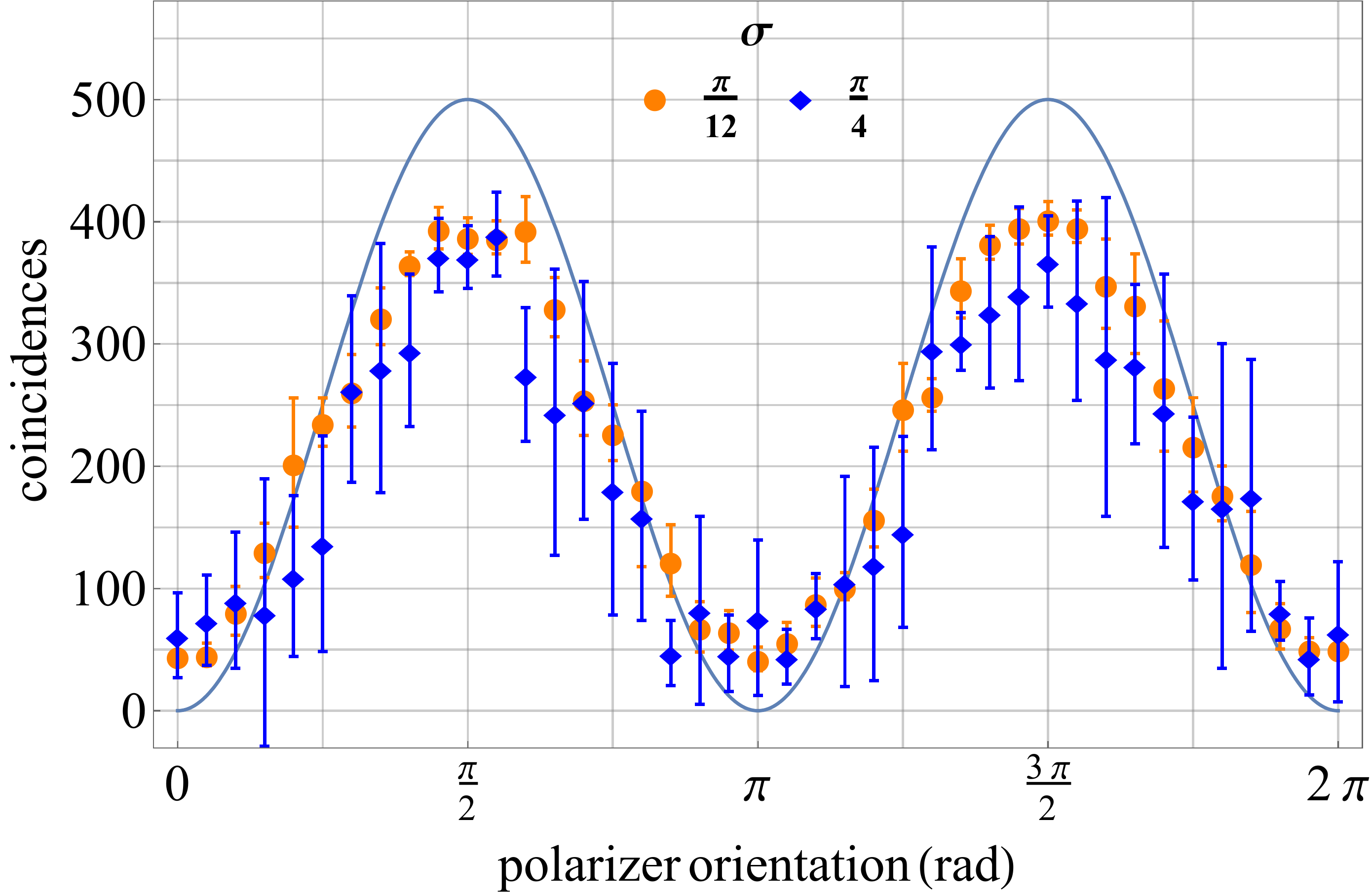}}\end{tabular}\\
\hline
			\centered{$100$}&\begin{tabular}{l}\raisebox{- \height}{\includegraphics[width=0.85\columnwidth]{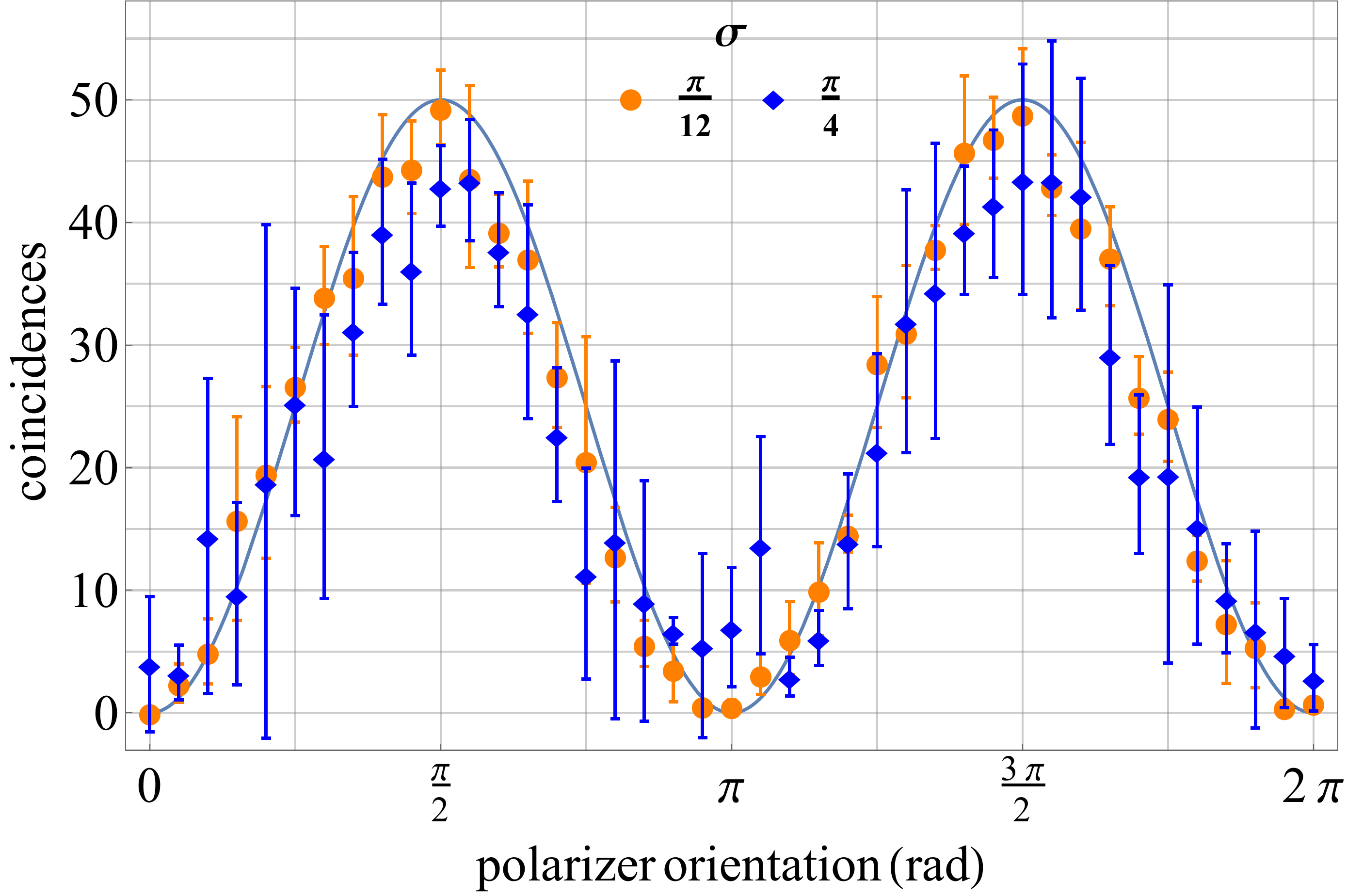}}\end{tabular}&
			\begin{tabular}{l}\raisebox{- \height}{\includegraphics[width=0.85\columnwidth]{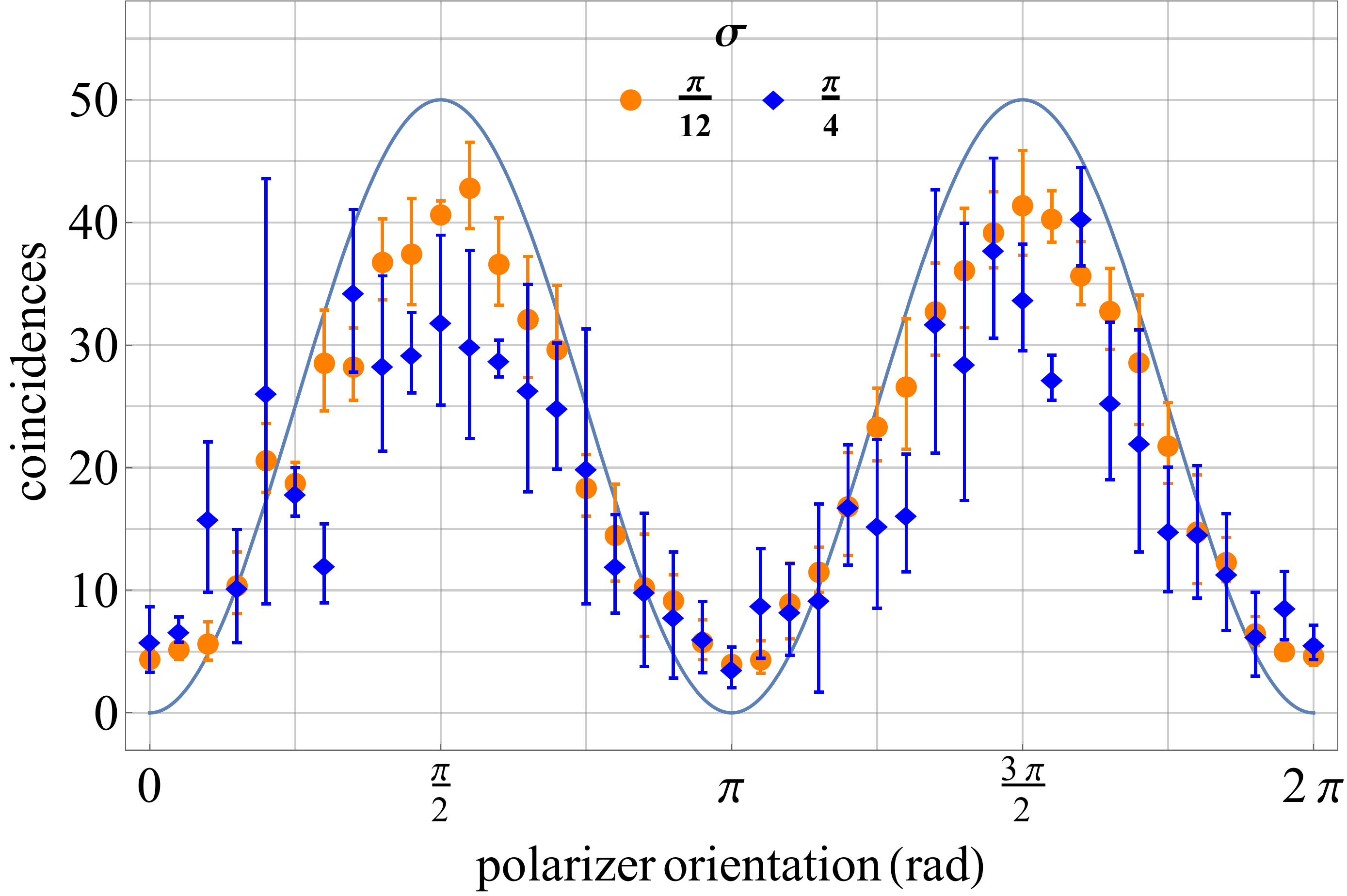}}\end{tabular}\\
\hline
			\centered{$10$}&\begin{tabular}{l}\raisebox{- \height}{\includegraphics[width=0.85\columnwidth]{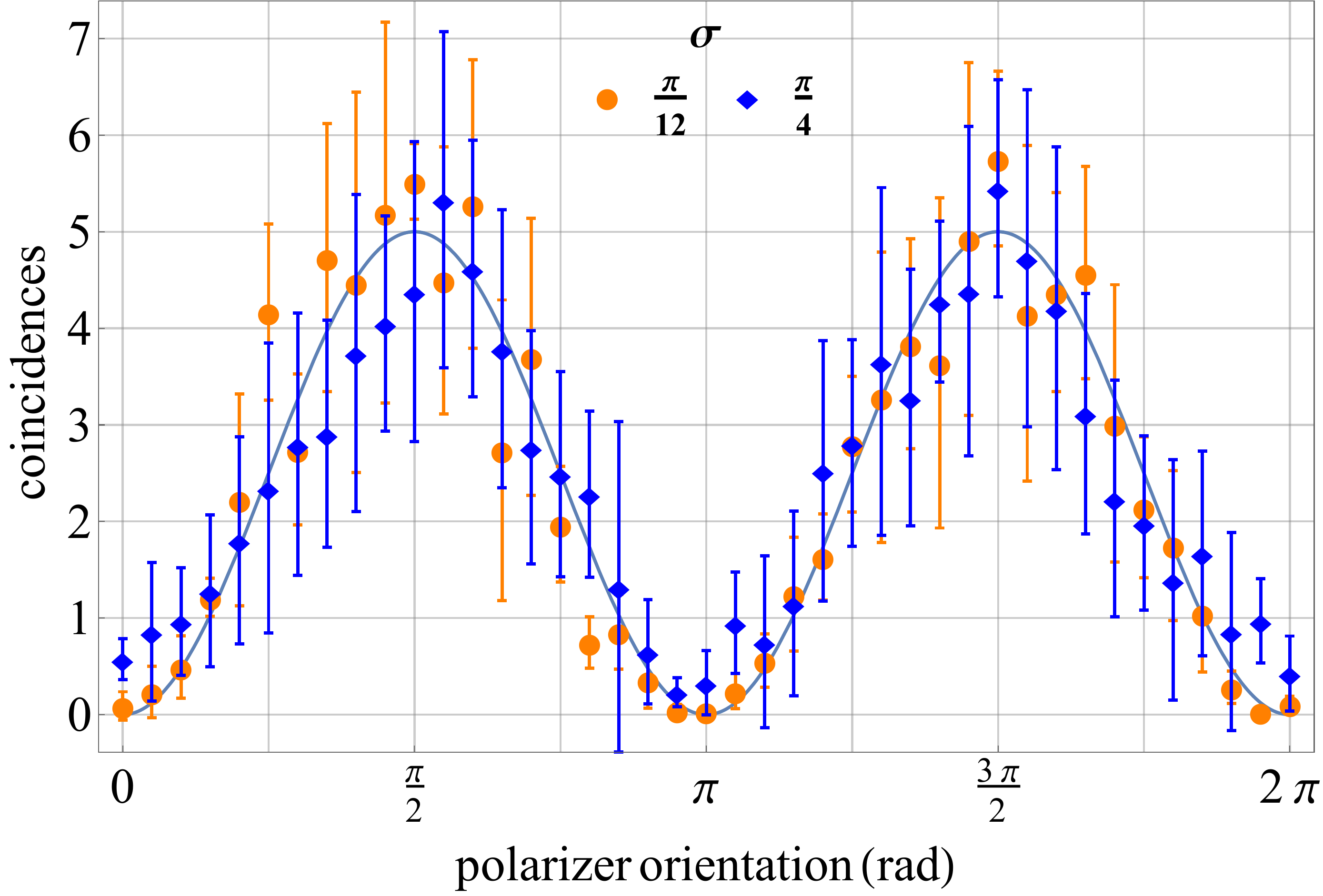}}\end{tabular}&
			\begin{tabular}{l}\raisebox{- \height}{\includegraphics[width=0.85\columnwidth]{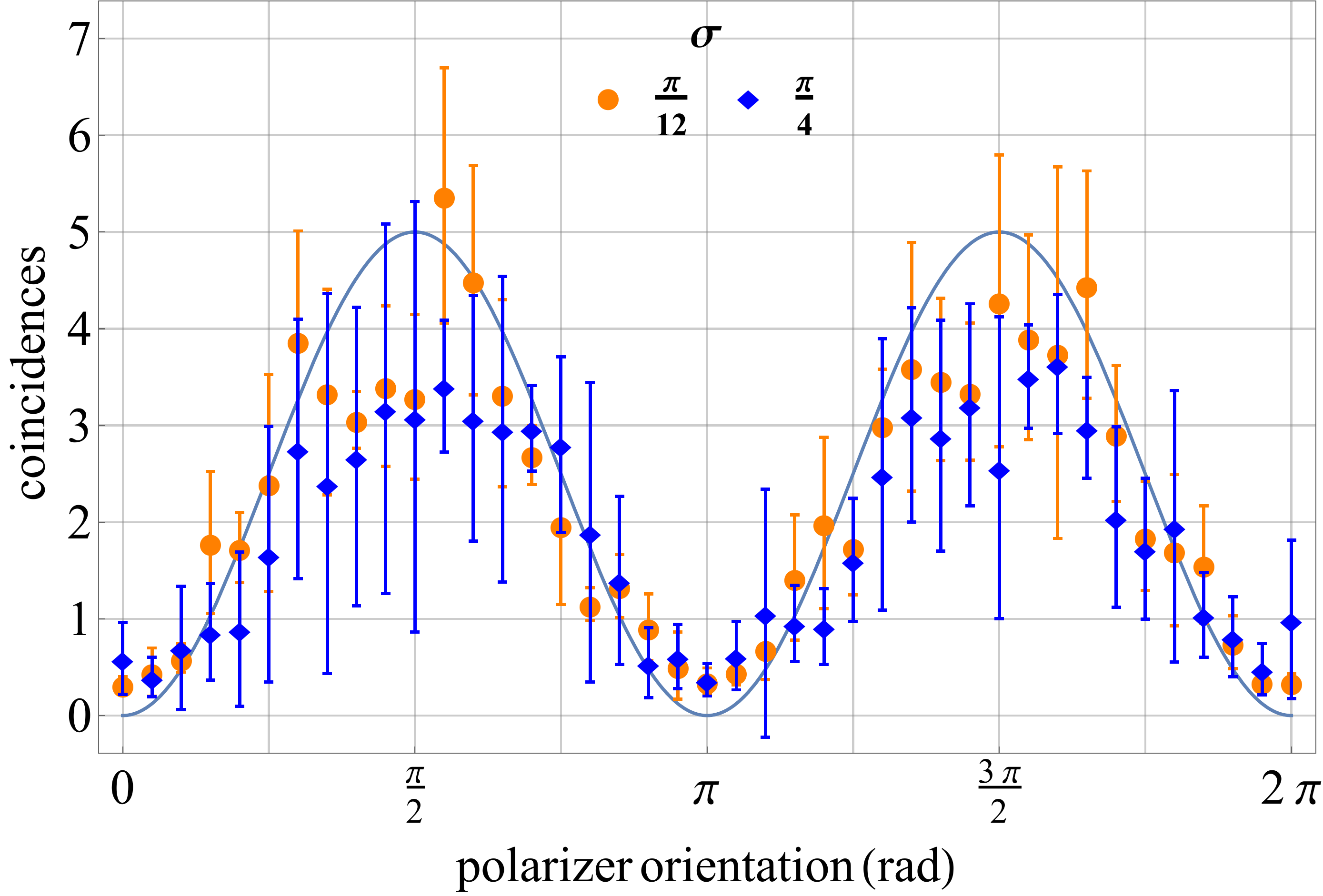}}\end{tabular}\\
\hline
		\end{tabular}
		\caption{Average pair count rates when the signal polarization analyzer measured horizontal polarization (H) while the other analyzer made many linear polarization measurements, for a fixed input state $\ket{\Phi^+} = \frac{1}{\sqrt{2}} \left(\ket{00} + \ket{11} \right)$. The zero angle corresponds to the vertical polarization (V) measurement. Error bars represent one standard deviation. }
		\label{correlations}
	\end{figure*}

First, numerical simulations were performed to investigate quantum correlations in a two-photon state with noisy measurements. The analysis of entanglement was realized by assuming that the polarization analyzer in one arm is fixed at horizontal (H) orientation, whereas the other analyzer rotates to make many linear polarization measurements. This approach is commonly implemented to study the quality of entanglement, see Ref.~\cite{Horn2013,Nomerotski2020}. In \figref{correlations}, we present average pair count rates (coincidences), assuming different scenarios, for a fixed input state $\ket{\Phi^+} = \frac{1}{\sqrt{2}} \left(\ket{00} + \ket{11} \right)$. All sources of experimental errors were taken into account. The theoretical line represents the average coincidence rate for an ideal scenario based on the formula: $n (\theta) = \mathcal{N} \,\tr \left( M(\theta) \ket{\Phi^+}\!\bra{\Phi^+}\right)$, where $M(\theta)$ denotes the measurement operator. We have $M(\theta)=\ket{H}\!\bra{H} \otimes \ket{\theta}\!\bra{\theta}$, where $\ket{H}$ is the vector of horizontal polarization and $\ket{\theta} = \begin{pmatrix} \sin \theta & \cos \theta \end{pmatrix}^T$ corresponds to different linear polarization measurements for $0 \leq \theta < 2\pi$.

In each subfigure, we have two plots that correspond to different values of $\sigma$, which quantifies the noise introduced by random unitary rotations. In this scheme, we assumed that the analyzer measuring H polarization is burdened with a fixed error described by a unitary operator of the form \eqref{m3}. In contrast, each measurement of the rotating analyzer is associated with a random error due to angular uncertainties. Two specific values of $\sigma$ were selected, i.e., $\sigma=\pi/12$ and $\sigma=\pi/4$, to study how quantum correlations depend on this kind of noise.One can notice that the results for $\sigma=\pi/4$ feature more statistical dispersion, which is presented as error bars. This implies that for a series of measurements we obtain results that are scattered along a wide interval, whereas for $\sigma=\pi/12$ the counts are more concentrated, which proves better precision of the measurement.

In the rows of \figref{correlations}, one finds the results for three different numbers of photon pairs. By comparing the plots in one column consecutively, we notice the influence of the Poisson noise, which introduces more variance into the results for a smaller number of photon pairs. Then, by comparing the two columns, we can analyze the effect of receiving by the detector a portion of maximally mixed states as defined in \eqref{m15}. For a given $\mathcal{N}$, if we contrast the plots in both columns, we see that the background noise, which reaches the detector, makes the results deflect from the theoretical line. In other words, the background noise reduces the amount of quantum correlations in the system.

\subsection{Entanglement detection and state tomography}

\begin{figure}[h]
	\centering
  \begin{subfigure}{1\columnwidth}
\caption{average fidelity}
       \centering
         \includegraphics[width=0.95\columnwidth]{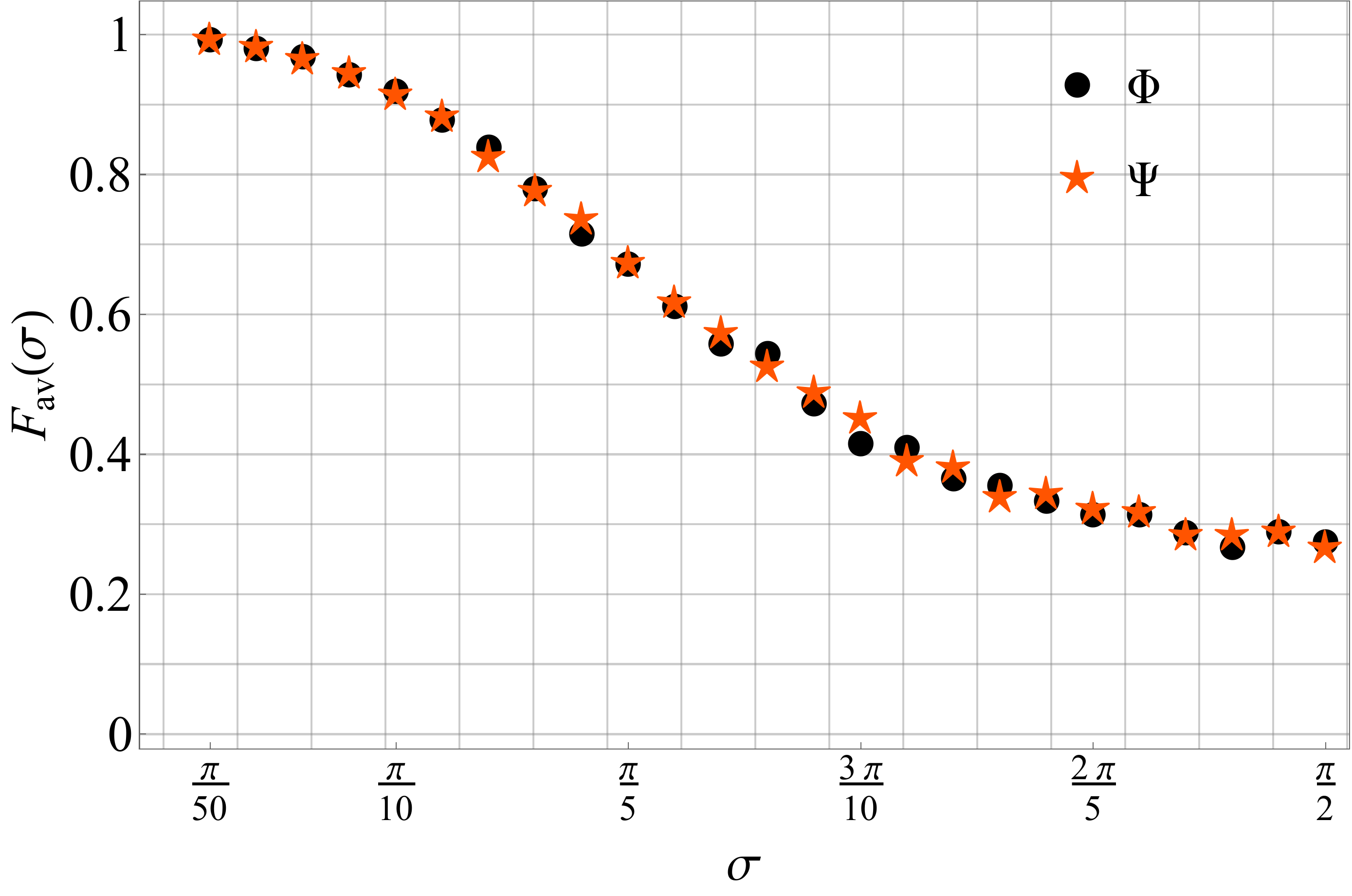}
\label{fidelity1}
     \end{subfigure}
     \hfill
     \begin{subfigure}{1\columnwidth}
\caption{average concurrence}
       \centering
         \includegraphics[width=0.95\columnwidth]{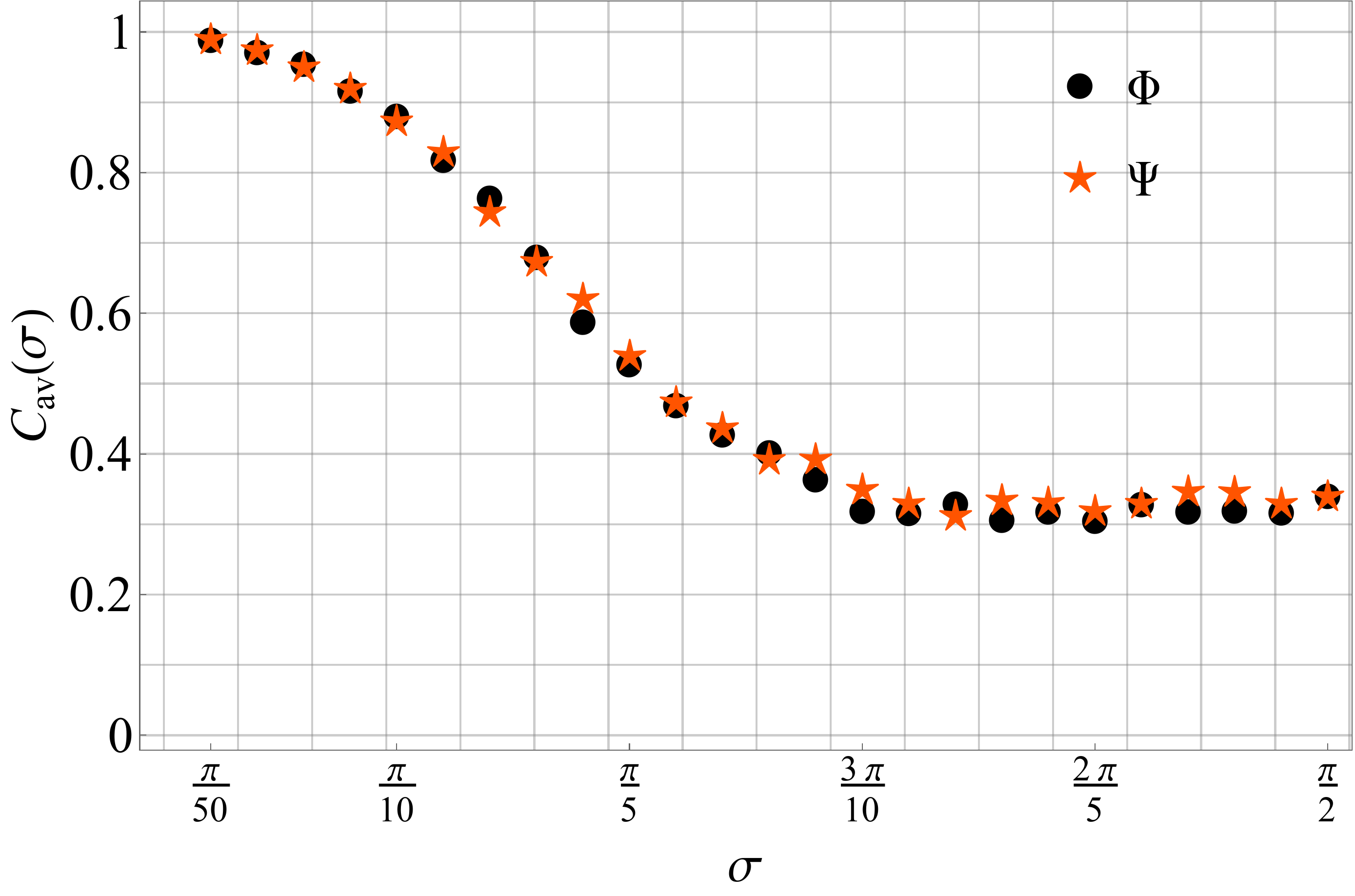}
\label{concurrence1}
     \end{subfigure}
	\caption{Plots present the average fidelity, $\mathcal{F}_{av} (\sigma)$, (the upper graph) and the concurrence $C_{av} (\sigma)$ (the lower graph) for QST of two families of entangled states. Each point was obtained for a sample of $200$ input states. The formula \eqref{m14} was applied with the noise governed by the standard deviation $\sigma$. The average number of photon pairs is fixed, $\mathcal{N} = 1\,000$.}
	\label{plots1}
\end{figure}

First, we intend to compare the efficiency of the QST framework for the two classes of maximally entangled states, i.e., $\Phi$ introduced in \eqref{m0} and $\Psi$ as in \eqref{m1}. For each family, we select a sample of $200$ input states such that the relative phase covers the full range. Every state is reconstructed in the scenario involving the Poisson noise and random unitary rotations, i.e. we follow the formula \eqref{m14} for the measured two-photon counts. The average fidelity (computed over the sample) is presented in \figref{fidelity1} as a function of $\sigma$ whereas \figref{concurrence1} presents the average concurrence.

In \figref{fidelity1}, one can observe that the accuracy of state reconstruction degenerates gradually as we increase the value of $\sigma$. It comes as no surprise since $\sigma$ quantifies the amount of experimental noise, and the quality of the QST framework was expected to decline along with this figure. In particular, for both $\Phi$ and $\Psi$, we have $\mathcal{F}_{av}(\pi/50)= 0.99 \pm 0.01$, which proves that the framework allows for accurate state reconstruction with minor noise. As we increase $\sigma$, the standard deviation for the average fidelity escalates. To be more specific, we obtain $\mathcal{F}_{av}(\pi/2)= 0.28 \pm 0.16$, which demonstrates the influence of the random noise on this figure of merit. Interestingly, there is no significant difference between the analyzed families of entangled states.

In \figref{concurrence1}, we see the amount of entanglement detected by our tomographic technique. At the beginning, we have $C_{av}(\pi/50)= 0.99 \pm 0.01$, which confirms that a minor amount of random noise does not reduce the ability to detect entanglement. Then, the function $C_{ac} (\sigma)$ decreases up to $\sigma = (3 \pi)/10$, when the value of the average concurrence stabilizes and remains equal approx. $0.3$. This means that the measurement technique is capable of detecting some entanglement in spite of experimental noise. At the end, we obtain $C_{av}(\pi/50)= 0.32 \pm 0.19$. The value of the standard deviation implies that the results for the sample feature a great deal of variation due to the random rotations. Just as $\mathcal{F}_{av} (\sigma)$, the average concurrence does not display any significant difference between the families $\Phi$ and $\Psi$.

\begin{figure}[h]
	\centering
  \begin{subfigure}{1\columnwidth}
\caption{average fidelity for the $\Phi-$class}
       \centering
         \includegraphics[width=0.95\columnwidth]{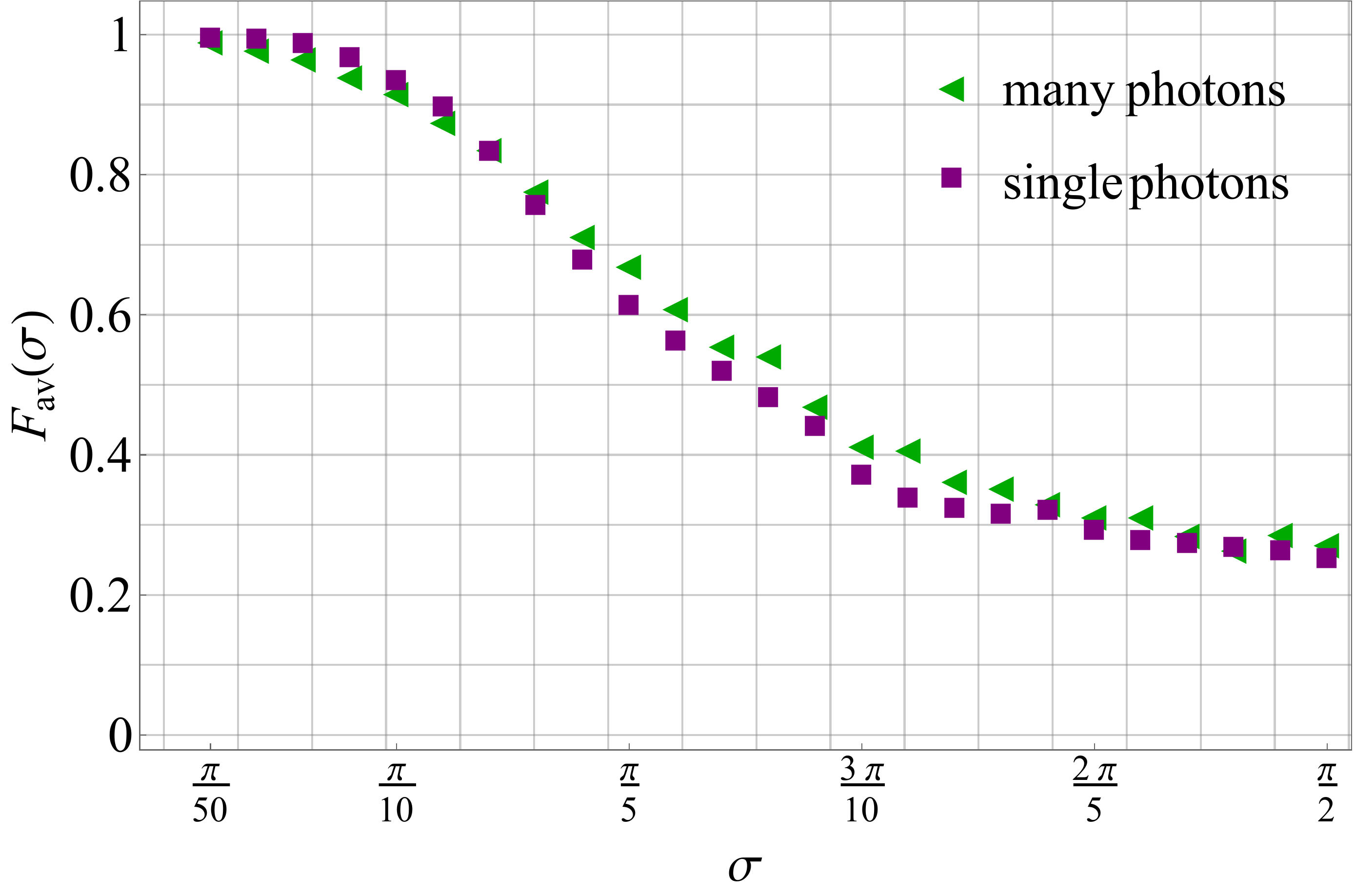}
\label{fidelity2}
     \end{subfigure}
     \hfill
     \begin{subfigure}{1\columnwidth}
\caption{average concurrence the $\Phi-$class}
       \centering
         \includegraphics[width=0.95\columnwidth]{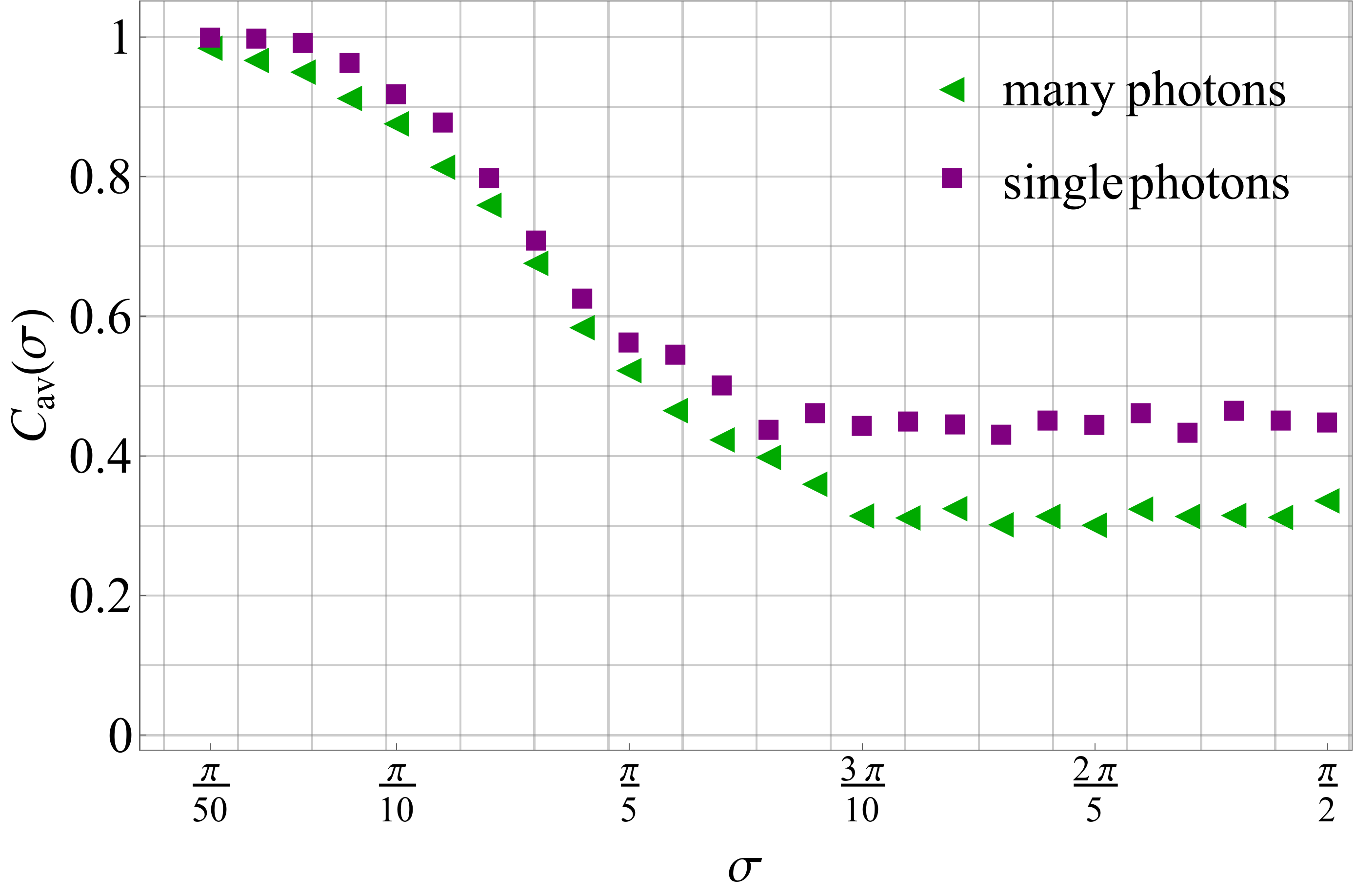}
\label{concurrence2}
     \end{subfigure}
	\caption{Plots present the average fidelity, $\mathcal{F}_{av} (\sigma)$, (the upper graph) and the concurrence $C_{av} (\sigma)$ (the lower graph) for QST of the $\Phi-$class. The average number of photon pairs is either $\mathcal{N} = 1\,000$ (many photons) or $\mathcal{N} = 10$ (single photons).}
	\label{plots2}
\end{figure}

The results presented in \figref{plots1} were obtained, assuming the average number of photon pairs equals $1\,000$. Therefore, it appears justified to examine how the accuracy of the method depends on the number of photon pairs involved in measurements. Since there was no significant difference between the two families of entangled states, we consider only the $\Phi-$family \eqref{m0} and select a sample of $200$ states (for different values of the relative phase $\alpha$). Once more, we follow the formula \eqref{m14} for the measured coincidences (no dark counts) and evaluate the performance of the framework versus the amount of experimental noise, which is presented in \figref{plots2}.

We distinguish two scenarios -- many photons (with $\mathcal{N} = 1\, 000$) and single photons (assuming $\mathcal{N} = 10$). As far as the average fidelity is concerned, in \figref{fidelity2}, we can observe that both plots present the same tendency. There is no significant difference in the accuracy of quantum state estimation due to distinct numbers of photon pairs involved in measurements. However, one may notice a modest advantage of the many-photon approach for the middle values of $\sigma$. This effect is not considerable, though. Both scenarios feature a very similar value of the standard deviation corresponding to each average fidelity, which was not plotted for the sake of the figure's clarity.

In \figref{concurrence2}, we observe the average concurrence versus the amount of experimental noise. These results suggest that the scenario with single photons outperforms the many-photon case when it comes to entanglement detection. In particular, for greater amounts of noise, i.e. $\sigma \geq (7 \pi)/25$, we see that the average concurrence for the single-photon scenario satisfies: $0.4 < C_{av} (\sigma) < 0.5$. However, to better understand this effect, we need to compare the value of standard deviation, which greater in the case of $\mathcal{N} = 10$, e.g., $C_{av} (\pi/2) = 0.52 \pm 0.24$. This implies that the single-photon scenario leads to a better average fidelity, but at the same time, it involves more variance.

\begin{figure}[h]
	\centering
\textbf{Many-photon scenario}\par\medskip
  \begin{subfigure}{1\columnwidth}
\caption{average fidelity for different dark count rates}
       \centering
         \includegraphics[width=0.95\columnwidth]{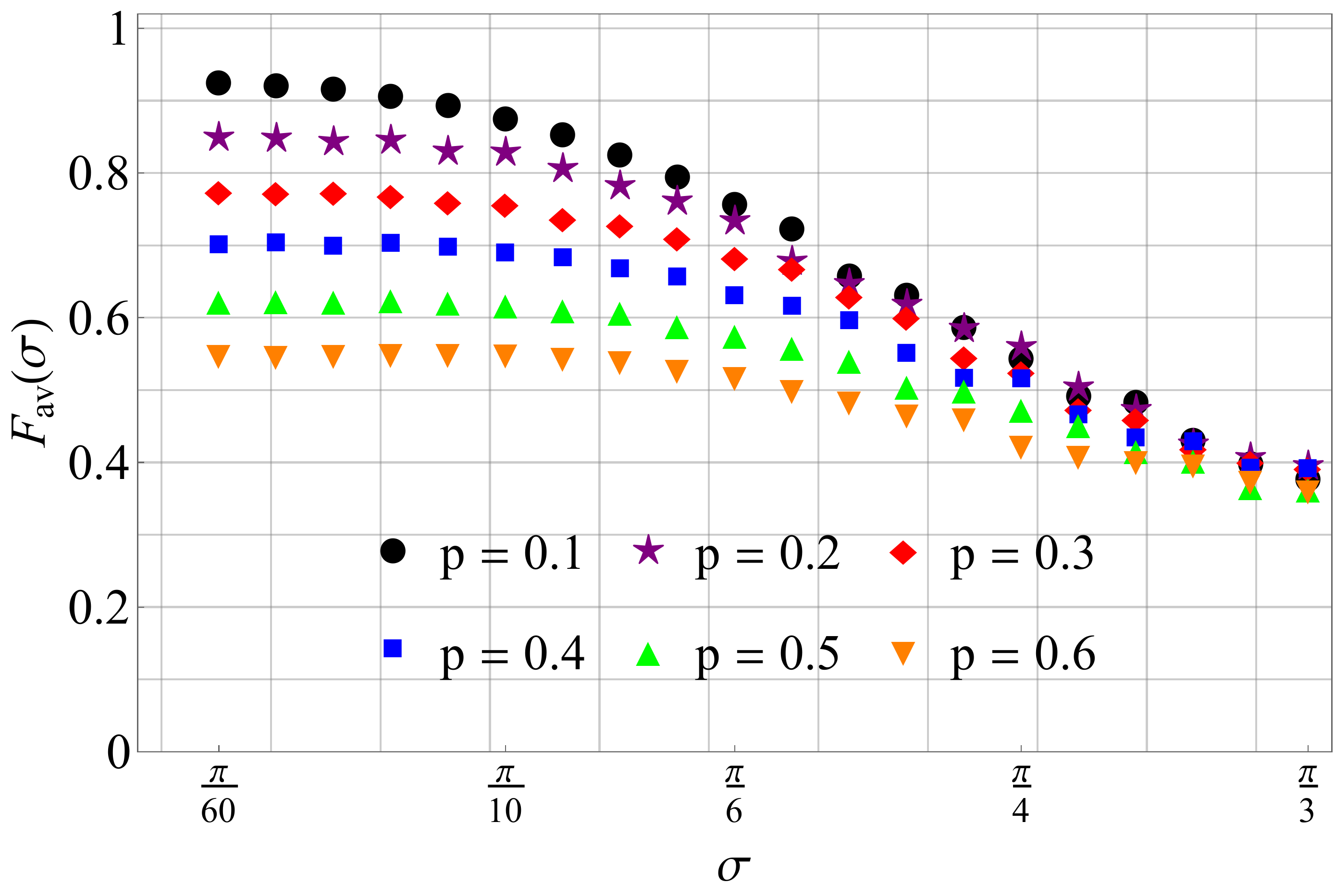}
\label{fidelity3}
     \end{subfigure}
     \hfill
     \begin{subfigure}{1\columnwidth}
\caption{average concurrence for different dark count rates}
       \centering
         \includegraphics[width=0.95\columnwidth]{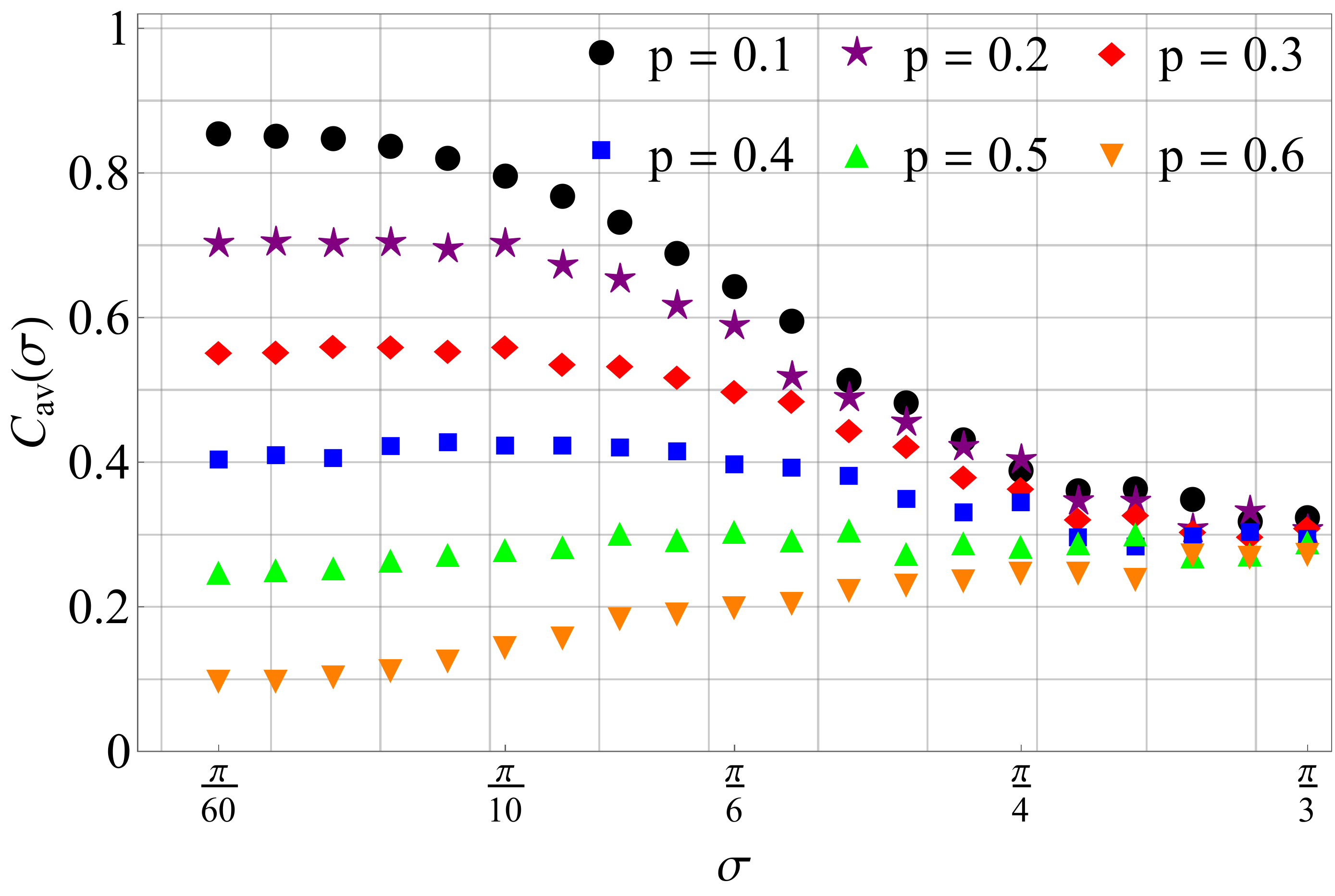}
\label{concurrence3}
     \end{subfigure}
	\caption{Plots present the average fidelity $\mathcal{F}_{av} (\sigma)$ and the concurrence $C_{av} (\sigma)$ for QST of the $\Phi-$family with selected dark count rates. The average number of photon pairs is: $\mathcal{N} = 1\,000$.}
	\label{plots3}
\end{figure}

The advantage of the single-photon scenario in entanglement detection appears intriguing since, for a low number of photons, one would expect the Poisson noise to have a more detrimental impact on the accuracy of QST. Nevertheless, for most applications, we are interested in detecting such entangled states that are sufficient to announce the violation of the CHSH inequality \cite{Clauser1969}, which is a generalization of the original Bell's inequality \cite{Bell1964}. Based on the concurrence, we can conclude that the CHSH inequality is violated if $C[\rho] > 1/\sqrt{2}$ \cite{Verstraete2002,Hu2012}. In our application, it means that the sufficient condition for the violation of the CHSH inequality can be expressed as: $\sigma < (4 \pi)/25$. For such values of $\sigma$, the single-photon scenario has only a modest advantage over the many-photon case.

It is worth noting that the number of photon pairs in the many-photon scenario may take different values since numerical simulations have confirmed that the plot of $C_{av} (\sigma)$ remains unchanged if $\mathcal{N} = 3\,000$ or $\mathcal{N} = 5\,000$.

In the next step, the QST framework is tested by adding dark counts, see \eqref{m15}, where the parameter $p$ can be referred to as the \textit{dark count rate}. Again, we consider only the $\Phi-$family of two-qubit entangled states \eqref{m0} and we examine the accuracy of the framework for different non-zero values of the dark count rate, i.e. $p \in \{0.1, 0.2, \dots, 0.6\}$.

\begin{figure}[h]
	\centering
\textbf{Single-photon scenario}\par\medskip
  \begin{subfigure}{1\columnwidth}
\caption{average fidelity for different dark count rates}
       \centering
         \includegraphics[width=0.95\columnwidth]{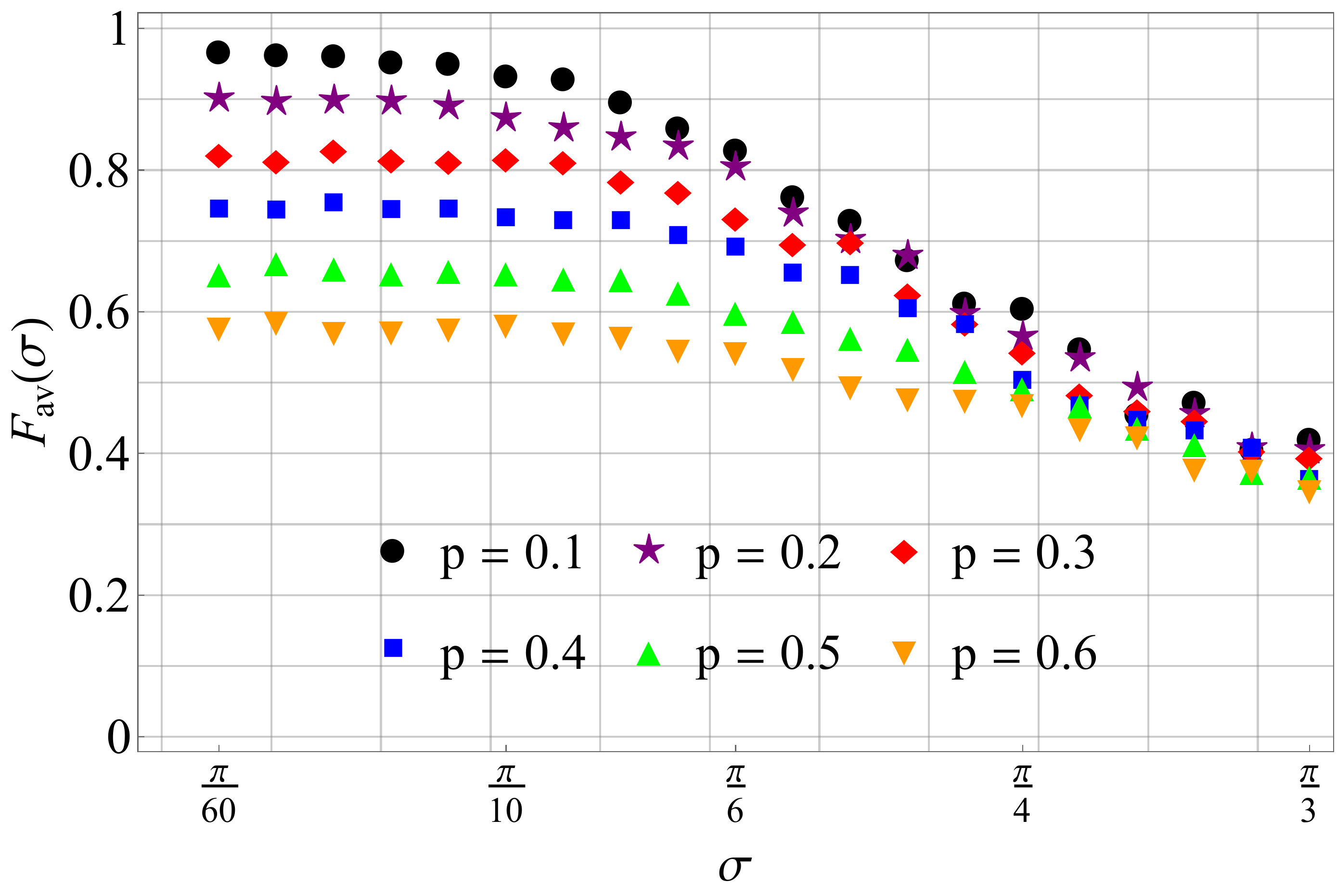}
\label{fidelity4}
     \end{subfigure}
     \hfill
     \begin{subfigure}{1\columnwidth}
\caption{average concurrence for different dark count rates}
       \centering
         \includegraphics[width=0.95\columnwidth]{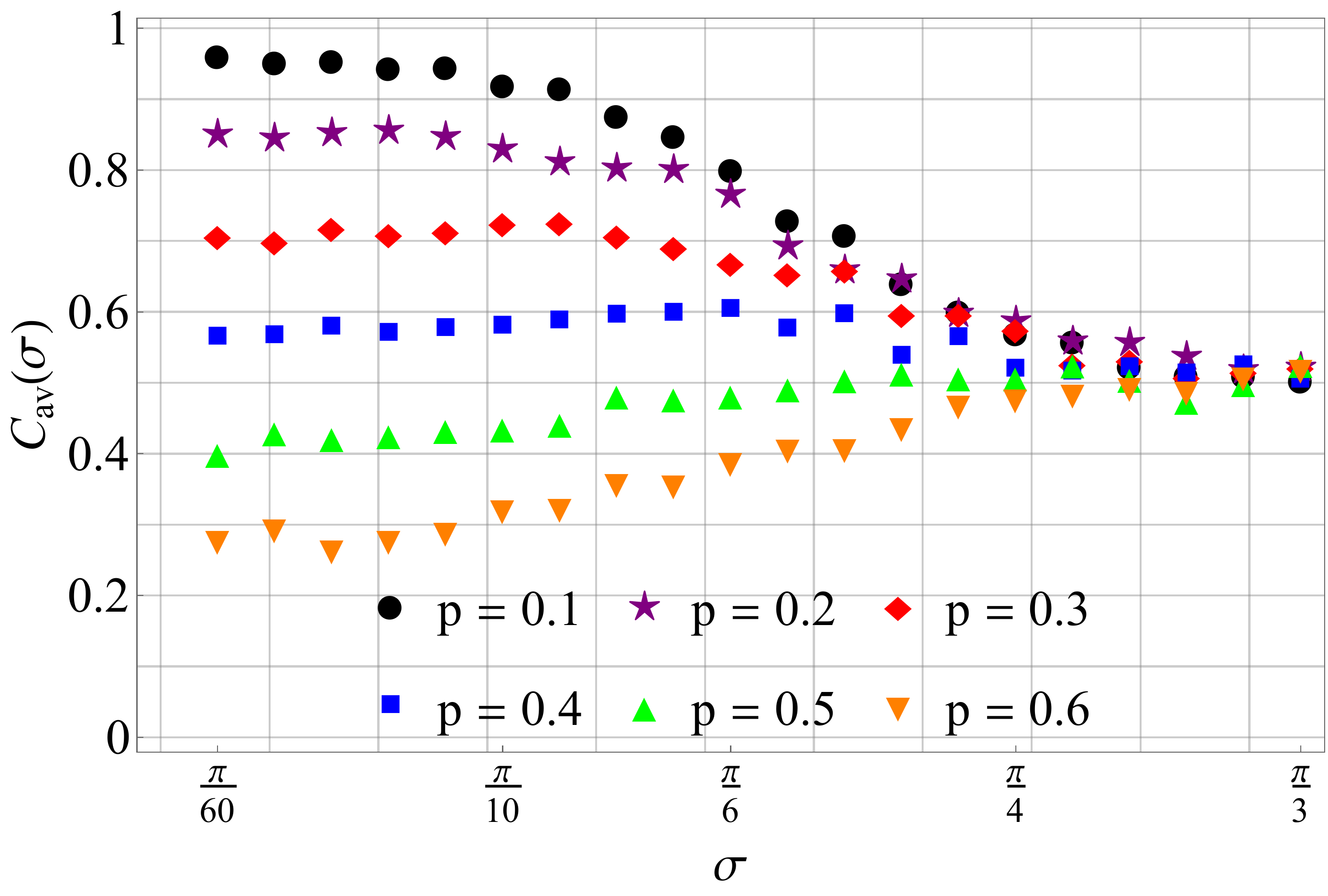}
\label{concurrence4}
     \end{subfigure}
	\caption{Plots present the average fidelity $\mathcal{F}_{av} (\sigma)$ and the concurrence $C_{av} (\sigma)$ for QST of the $\Phi-$family with selected dark count rates. The average number of photon pairs is: $\mathcal{N} = 10$.}
	\label{plots4}
\end{figure}

In \figref{plots3}, one can find the plots which were obtained, assuming that the average number of photon pairs is: $\mathcal{N} = 1\,000$ (many-photon scenario). For selected dark count rates, we can observe the properties of $\mathcal{F}_{av} (\sigma)$ and $C_{av} (\sigma)$ versus the amount of noise introduced by unitary rotations, which is quantified by $\sigma$. Interestingly, all plots converge as we increase $\sigma$, which means that for a great amount of noise due to unitary rotations adding dark counts does not change the accuracy of the framework. However, when $\sigma$ is small, the impact of dark counts on entanglement detection appears very detrimental. In particular, for $\sigma = \pi/60$, we observe that increasing the dark count rate by $0.1$ generates a decline in the average concurrence approximately equal to $0.15$. Finally, we can notice that if $p \geq 0.2$ we are not able to guarantee the violation of the CHSH inequality since for such dark count rates we have $C_{av} (\sigma) < 1/\sqrt{2}$ irrespective of $\sigma$.

Next, we consider the same kind of problem in the single-photon scenario, i.e., the average number of photon pairs is reduced: $\mathcal{N} = 10$. The figures of merit, for the same set of dark count rates, are given in \figref{plots4}. By comparing \figref{fidelity3} and \figref{fidelity4}, we see that in principle, both models demonstrate very close accuracy as far as the average fidelity is concerned (though a modest lead of the single-photon scenario is noticeable).

\begin{figure}[h]
	\centering
\textbf{Single-photon versus many-photon scenario}\par\medskip
  \begin{subfigure}{1\columnwidth}
\caption{average fidelity for different dark count rates}
       \centering
         \includegraphics[width=0.95\columnwidth]{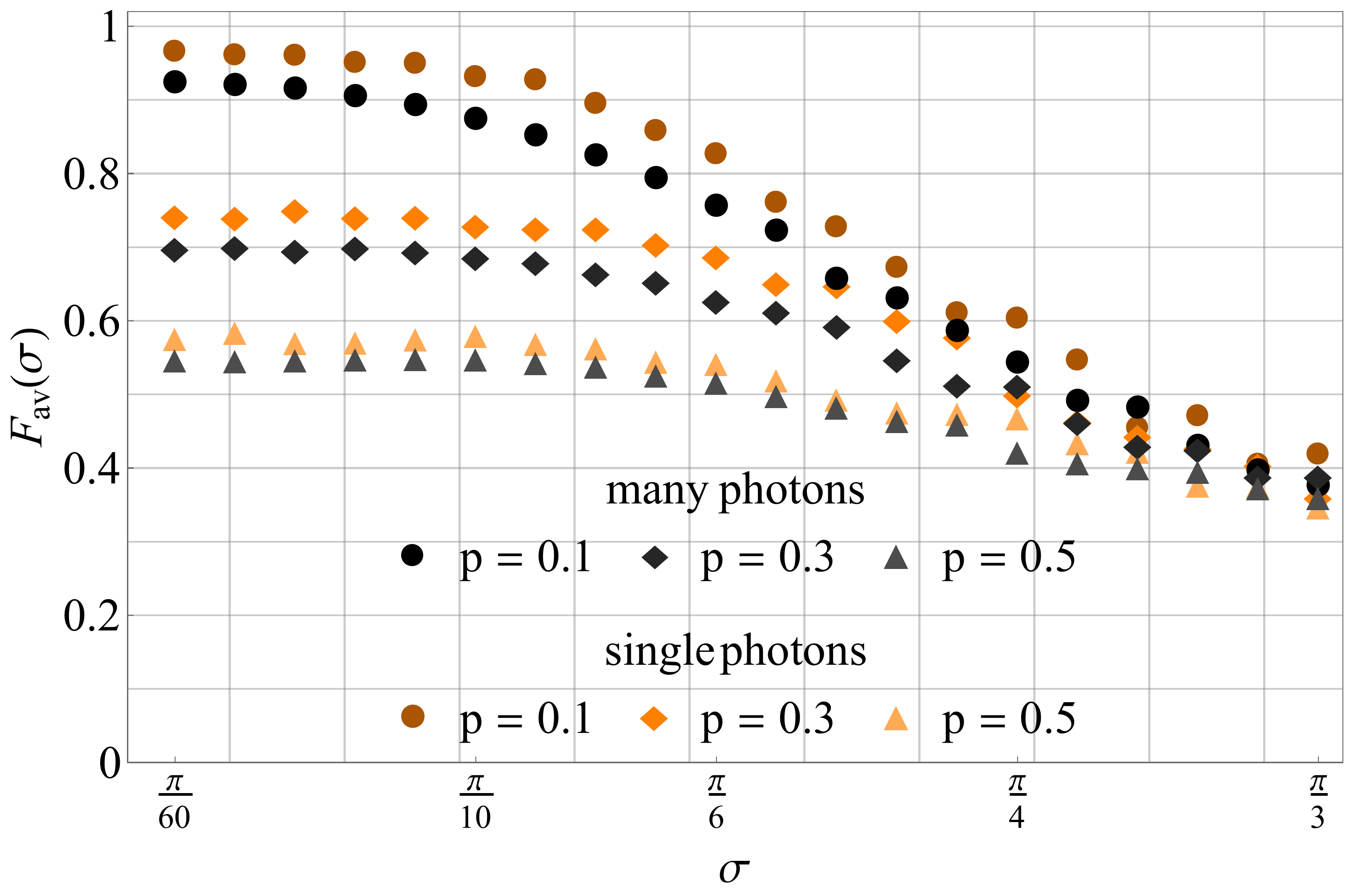}
\label{fidelity5}
     \end{subfigure}
     \hfill
     \begin{subfigure}{1\columnwidth}
\caption{average concurrence for different dark count rates}
       \centering
         \includegraphics[width=0.95\columnwidth]{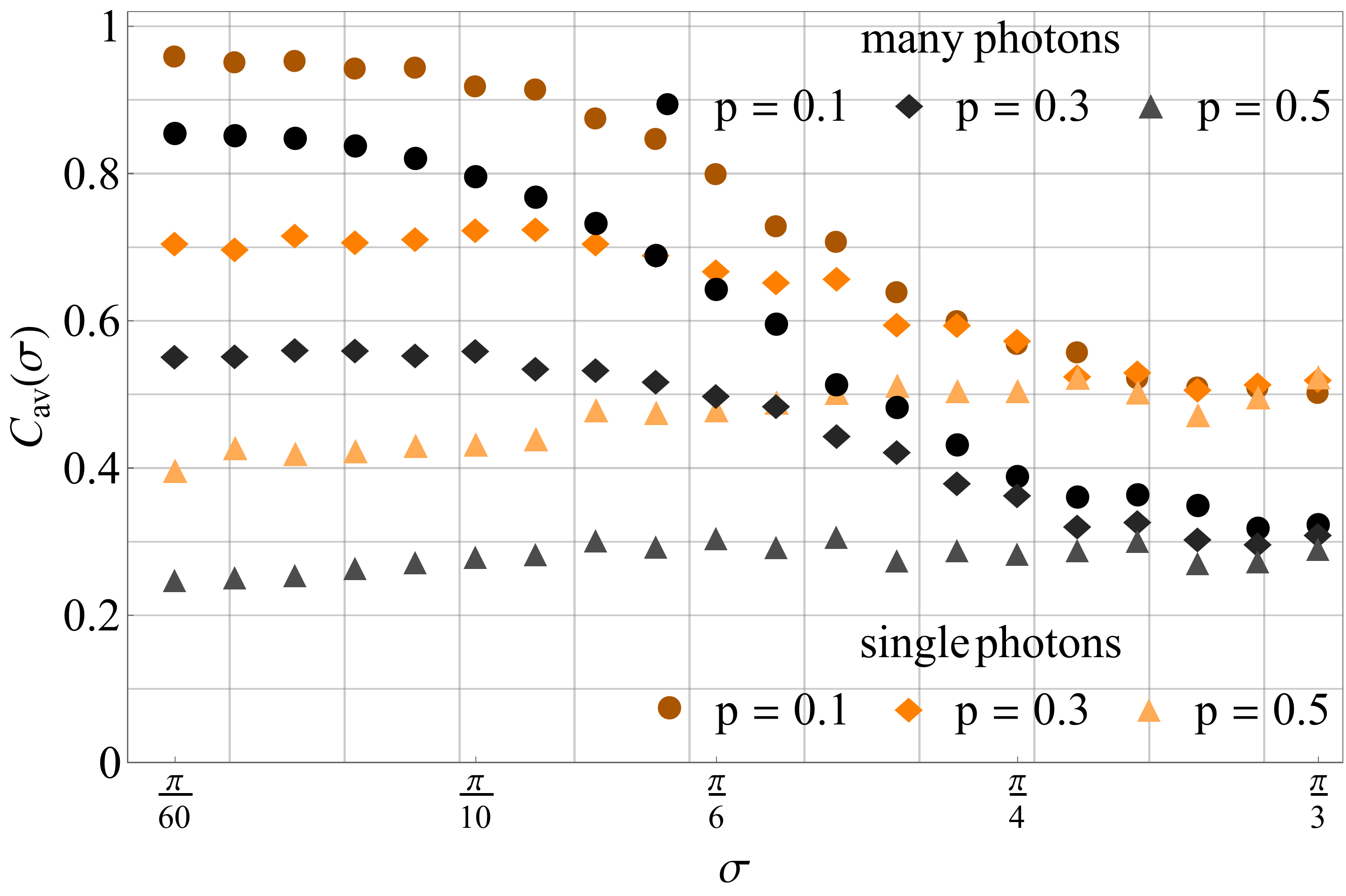}
\label{concurrence5}
     \end{subfigure}
	\caption{Plots present the average fidelity $\mathcal{F}_{av} (\sigma)$ and the concurrence $C_{av} (\sigma)$ for QST of the $\Phi-$family with selected dark count rates. Two scenarios with different number of photon pairs are compared.}
	\label{plots5}
\end{figure}

Furthermore, from \figref{concurrence3} and \figref{concurrence4}, we can conclude that there is a substantial difference between the scenarios in entanglement detection. For any dark count rate, the single-photon scenario leads to a better average concurrence. If we utilize many photon pairs per measurement, then all plots of $C_{av} (\sigma)$ converge at approximately $0.3$, which is evident from \figref{concurrence3}. However, by following the single-photon scenario, we obtain plots of the average concurrence that for greater values of $\sigma$ converge at $0.5$. Nevertheless, it should be stressed that the sample in the single-photon scenario features a greater deal of variance.

To demonstrate more explicitly the advantage of the single-photon scenario over the many-photon approach as long as the average values are concerned, we arrange the plots of $\mathcal{F}_{av} (\sigma)$ and $C_{av} (\sigma)$ for both models in \figref{plots5}. For the clarity of the presentation, only three dark count rates are considered. The plots in \figref{fidelity5} confirm that the single-photon scenario slightly outperforms the other model when we take into account the average fidelity. It means that with single photons, we obtain estimates that, on average, more accurately represent actual input states.

In \figref{concurrence5}, one can track how all three sources of errors (i.e., the Poisson noise, unitary rotations, and dark counts) influence the detection of entanglement. We see that for each dark count rate, the plot corresponding to the single-photon scenario lies above the corresponding graph relating to the many-photon approach. These results demonstrate that the single-photon scenario is more efficient in entanglement detection than the other approach.

\section{Discussion and summary}\label{discussion}

In the article, we introduced a framework for QST of entangled qubits based on noisy measurements, which are distorted by the Poisson noise, dark counts, and unitary rotations. Initially, the framework was tested on two classes of entangled states. Since no difference was found in the accuracy of the framework, further investigation was devoted to the $\Phi-$family of entangled qubits because this class is famous for practical implementations, in particular for \textit{quantum key distribution} (QKD).

First, we considered a model without dark counts with different numbers of photon pairs involved in each measurement. The results allowed us to observe that the single-photon scenario slightly outperforms the many-photon approach in terms of entanglement detection quantified by the average concurrence, especially for greater amounts of errors due to unitary rotations, as presented in \figref{concurrence2}. This result may appear intriguing as one would expect that by utilizing a greater number of photon pairs, we should obtain better performance of the QST technique. However, the concurrence computed for a sample in the single-photon scenario features more variance, which diminishes the advantage.

Then, the model was extended by including dark counts. For a set of dark count rates, we can observe in detail how the efficiency of the framework changes as we increase $\sigma$. The single-photon scenario appears more advantageous for any non-zero dark count rate in terms of both reconstruction accuracy and entanglement detection. It can be explained by noticing that the presence of dark counts multiplies errors more significantly for a higher number of photon pairs. From \eqref{m16}, one can see that if we increase the number of photon pairs per measurement, we reduce the Poisson noise, but on the other hand, it magnifies the influence of the dark count rate. Thus, taking all sources of noise into account, the single-photon scenario seems optimal.

The findings described in the article provoke new scientific questions. It requires more research to determine why the single-photon scenario outperforms the many-photon approach in entanglement detection without dark counts. Next, it should be examined whether the same tendency takes place if we consider different noise models.

\section*{Acknowledgments}

The author acknowledges financial support from the Foundation for Polish Science (FNP) (project First Team co-financed by the European Union under the European Regional Development Fund).

\end{document}